\documentclass[manuscript]{aastex}

\usepackage{color}

\shorttitle{}
\shortauthors{Monje et al.}

\begin{document}


\title{Discovery of Hydrogen Fluoride in the Cloverleaf Quasar at z = 2.56}


\author{R. R. Monje, T. G. Phillips} 
\affil{California Institute of Technology, 1200 E. California Blvd., Pasadena, CA  91125-4700, USA}
\email{raquel@caltech.edu}
\author{R. Peng} 
\affil{Caltech Submillimeter Observatory, 111 Nowelo Street, Hilo, HI 96720, USA}
\author{D. C. Lis}
\affil{California Institute of Technology, 1200 E. California Blvd., Pasadena, CA  91125-4700, USA}
\author{D. A. Neufeld} 
\affil{Department of Physics and Astronomy, Johns Hopkins University, 3400 North Charles Street, Baltimore, MD 21218, USA}
\and
\author{M. Emprechtinger}
\affil{California Institute of Technology, 1200 E. California Blvd., Pasadena, CA  91125-4700, USA}




\begin{abstract}

We report the first detection of hydrogen fluoride (HF) toward a high redshift quasar. Using the Caltech Submillimeter Observatory (CSO) we detect the HF $J=1-0$ transition in absorption toward the Cloverleaf, a broad absorption line (BAL) quasi-stellar object (QSO) at z=2.56. The detection is statistically significant at the $\sim$ 6$\sigma$ level. We estimate a lower limit of 4 $\times$ 10$^{14}$ cm$^{-2}$ for the HF column density and using a previous estimate of the hydrogen column density, we obtain a lower limit of 1.7 $\times$ 10$^{-9}$ for the HF abundance. This value suggests that, assuming a Galactic N(HF)/N$_H$ ratio, HF accounts for at least $\sim$10\% of the fluorine in the gas phase along the line of sight to the Cloverleaf quasar. This observation corroborates the prediction that HF should be a good probe of the molecular gas at high redshift. Measurements of the HF abundance as a function of redshift are urgently needed to better constrain the fluorine nucleosynthesis mechanism(s).

\end{abstract}


\section{Introduction}
The astrochemistry of the early Universe is crucial for understanding the properties of molecular gas in the early epoch of galaxy formation and for providing fundamental constraints on galaxy evolution. To date, a fairly small number of species, other than CO, have been detected at z $>$ 2 \citep[e.g. HCN, HNC, CN, H$_2$O;][]{Gue07, Rie10, Lis11, Omo11}. Observations with the \textit{Herschel Space Observatory} \citep{Pil10} have allowed for the first time the detection of the $J=1-0$ transition of hydrogen fluoride (HF) at 1.232~THz in the local universe, and revealed the ubiquitous nature of this molecule in the interstellar medium (ISM) of the Milky Way. Indeed, HF has proven to be a good tracer of molecular gas in the ISM, being detected in environments as diverse as Orion~KL, OMC-1 \citep{Phi10}, as well as diffuse clouds on the lines-of-sight toward W49N, W51 \citep{Son10}, W31C \citep{Neu10} and Sagittarius B2(M) \citep{Mon11} and in two nearby active galaxies, Mrk 231 \citep{van10} and Arp220 \citep{Ran11}. This transition is generally observed in absorption, as expected, due to its very large A--coefficient, A$_{10}$ = 2.42 $\times$ 10$^{-2}$~s$^{-1}$. Only an extremely dense region, with a strong radiation field, could generate enough excitation to yield an HF feature in emission \citep{Neu09}. 

We expect the HF to be the main reservoir of fluorine in the interstellar medium (ISM) because of its unique thermochemistry. Fluorine reacts exothermically with molecular hydrogen (H$_2$) to create a diatomic hydride molecule HF, forming a strong bond, only destroyed very slowly by means of photodissociation and as a result of reactions with ions of low abundances such as He$^+$, H$^{+}_{3}$, and C$^+$. HF absorption thus traces the total molecular column density along the line-of-sight, including very low column density or cold regions that may not be detectable in CO emission or other commonly used tracers of molecular hydrogen. Indeed, HF observations of Galactic sources have revealed absorption components that had previously been undetected in observations of any other molecules \citep[see Figure 2 in][]{Son10}.

Following-up on the Herschel discovery of the ubiquitous nature of HF, \cite{Lis11} searched for HF toward the ultra-luminous lensed starburst galaxy APM 08279+5255 at z=3.911. Their results led to the fortuitous detection of water at the highest-redshift to date, but a non-detection of the HF $J=1-0$ absorption feature, which may be related to the specific face-on geometry of this source. Because of its strong submillimeter continuum, the Cloverleaf galaxy at z = 2.558 is another interesting high-redshift candidate in which to search for HF.

The Cloverleaf, H1413+117 (14$^h$15$^m$46$^s$.23 11$^\circ$29$\arcmin$44$\arcsec$.0, J2000.0), a broad absorption line (BAL) quasi-stellar object (QSO), is among the most luminous galaxies yet detected in the universe \citep{Bar02}. It was identified as a lensed object with four bright image components with angular separation from 0.77$\arcsec$ to 1.36$\arcsec$ \citep[][]{Mag88}. \cite{Ven03} derived a lens model for the Cloverleaf using CO $J=7-6$ data obtained from the IRAM Plateau de Bure interferometer (PdBI). Their model suggests that the CO emission originates from a region of $\sim$ 800~pc. The large size of the CO source suggests that the Cloverleaf is a composite object, with a central powerful active galactic nucleus (AGN) and an extended starburst region similar to the nearby ultra luminous infrared galaxies (ULIRGs). The large luminosity, the substantial lense magnification and the composite nature of this object make the Cloverleaf a good candidate to perform absorption line studies toward the AGN, providing an excellent laboratory for studying gas evolution with redshift in galaxies. 

We have searched for the HF $J=1-0$ transition, redshifted to 346 GHz, in the Cloverleaf using the Caltech Submillimeter Observatory (CSO). These observations are described in Section 2 below. Our results are presented in Section 3 and discussed in Section 4. The detection of hydrogen fluoride reported here marks the first discovery of this molecule at a redshift z $>$ 0.042.

\section{Observations}

We have used the 10.4 m CSO telescope to observe the $J=1-0$ transition of HF, with rest frequency of 1232.4762~GHz \citep{nol87}, toward the Cloverleaf quasar. The observations were carried out during two independent observing runs in March and June 2011, using the 345 GHz wideband receiver with single sideband (SSB) system noise temperatures from 300 to 600~K. The receiver was tuned to the HF line, redshifted to the frequency of 346.396 GHz. The antenna full width half maximum (FWHM) beam size at this frequency is about 21$\arcsec$, with a beam efficiency 0.75. Pointing was checked regularly on planets and bright CO stars with typical pointing errors less than 5$\arcsec$ during one night. 
We used the 1 GHz bandwidth fast fourier transform spectrometer (FFTS) back-end with 8192 channels and the 1 GHz bandwidth acousto-optical spectrometer (AOS) with 2048 channels.   

The average spectra obtained with the low- and high-resolution backends after an on-source integration time of $\sim$ 4 hours show an absorption feature (Figure 1) with an integrated absorption strengths of 0.194 $\pm$ 0.034 and 0.202 $\pm$ 0.034 K km/s, respectively, corresponding to detections at the 5.7 and 5.9 sigma levels. The March and June data sets and the two backends are entirely consistent, each data set giving a statistically significant detection (following the conventional detection criterion: signal-to-noise ratio $>$ 3). 

Whereas the absorption depth is the same for both spectrometers, the continuum level is different, due to backend related instabilities. Assuming a high optical depth (i.e. saturated absorption), a good approximation given that the HF optical depths are typically larger than those of other submillimeter transitions, such as H$_2$O and CH, we estimate the SSB continuum level to be equal to the absorption depth and obtain an upper limit to the continuum level of 3.5 $\pm$ 1 mK which corresponds to 114 $\pm$ 33 mJy (using a conversion Jansky to Kelvin of S/T$_{mb}$ $=$ 32.5 J$_y$/K). This value, within the uncertainties, agrees with the $\sim$~70 mJy value obtained by extrapolating to the observed frequency the best-fit continuum of \cite{Bra09}.

For data reduction and analysis, the IRAM GILDAS package was used. Individual spectra were averaged by weighting by the noise in each spectrum. The final r.m.s. in both backends is consistent and has a value of 0.99 mK over 20 km s$^{-1}$.

\section{Results}

 Figure 2 shows the average spectrum obtained with the high resolution FFTS backend, smoothed to a 20 km s$^{-1}$ resolution, with a linear baseline subtracted. We apply a Gaussian fit to the observed spectrum to calculate the significance of the detected line. We obtain an absorption-line equivalent width of 55 km s$^{-1}$, a peak intensity of -4.1 mK and a central velocity of -113 km s$^{-1}$. The velocity shift is well within the width of the observed CO line profiles \citep{All97}. The equivalent width of the line is only twice our velocity resolution, therefore we cannot exclude the possibility that the line could have partially unresolved substructures. 

\cite{Rie11} found from their HCO$^+$ ($J=4-3$) excitation modeling that the dense gas excitation in the Cloverleaf is consistent with being purely collisional, rather than being enhanced by radiative processes, as suggested for similar sources like APM 08279+5255 (z = 3.91) where high-$J$ HCN transitions are substantially enhanced by radiative excitation through pumping of mid-infrared rovibrational lines \citep{Wei07,Wag05,Rie10}. The strong radiation field could have led to photodissociation of HF and explain its low abundance in this source \citep{Lis11}. Therefore, due to its large spontaneous emission coefficient (2.42 $\times$ 10$^{-2}$  s$^{-1}$) and low rates of collisional excitation, we expect that most HF molecules toward the Cloverleaf will be in the ground rotational state. In that case, we can estimate a lower limit for the total HF column density by use of the following expression:
\begin{equation}
	N_{tot} = \frac{8\pi^{3/2}\cdot\Delta\nu}{2\sqrt{ln2}A_{ul}\lambda^3}\cdot\frac{g_l}{g_u}\cdot\tau
\end{equation}

where $\Delta\nu$ is the line FWHM, $\lambda~=~243.2~\mu m$ is the wavelength of the observed transition, $A_{ul}~=~2.42\times 10^{-2}~s^{-1}$ is the Einstein coefficient, $g_l = 1$ and $g_u = 3$ are the degeneracy of the lower and upper levels and $\tau$ is the optical depth. Following the assumption of saturated absorption we use a $\tau$ greater than 3 for our lower limit calculations. We obtain a conservative lower limit on the total HF column density of $\sim$ 4 $\times$ 10$^{14}$ cm$^{-2}$. 

\section{Discussions and Conclusions}

The mechanism of fluorine nucleosynthesis still remains highly debated. To date, three different possible scenarios for the nucleosynthesis of fluorine have been proposed: neutrino capture by $^{20}$Ne in Type II supernovae proposed by \cite{Woo98}, production in AGB stars \citep[e.g.][]{Mow96} and in Wolf-Rayet stars \citep{Mey00}. However, the relative importance of these various processes remains unclear. Our detection of hydrogen fluoride towards the Cloverleaf quasar at z~=~2.558 represents the \textit{earliest} fluorine ever detected (just 2.6 Gyr after the big bang assuming H$_0$~=~71 km/s, $\lambda$~=~0.73, $\Omega$$_M$~=~0.27), and argues for significant production in high mass stars (i.e. Wolf-Rayet stars or supernovae).

From a chemical analysis of fluorine-containing molecules in the interstellar medium, \cite{Neu05} predicted that over a wide range of conditions the HF and H$_2$ column densities would track each other closely with the ratio of N(HF)/N(H$_2$) equal to the gas-phase elemental abundance of F relative to H nuclei \citep[equal to $\sim$ 3.6 $\times$ 10$^{-8}$ in diffuse atomic clouds in the solar neighborhood;][]{Sno07}. Herschel observations of HF $J=1-0$ toward a diverse Galactic environments proved that HF is an excellent probe of H$_2$ in diffuse clouds, although the derived abundances, with an average ratio of N(HF)/N(H$_2$)= 1.44 $\pm$ 0.35 $\times$ 10$^{-8}$, are a factor of two below the chemical model predictions \citep{Son10,Mon11}. Lower values of HF abundances have been obtained toward denser regions due to freeze out of HF onto dust grains \citep[e.g.][observed HF abundances as low as 5 $\times$ 10$^{-10}$ in the denser parts of NGC6334 I]{Emp11}. Our first detection of HF absorption line toward a high-redshift source shows that the HF line is also an extremely sensitive tool to trace molecular material toward high-redshift objects. From the HF absorption line observed toward the Cloverleaf we place a lower limit of 4 $\times$ 10$^{14}$ cm$^{-2}$ on the HF column density. \cite{Bra09} used the observed strength of the CO $J=6-5$, $J=8-7$ and $J=9-8$ lines and the disk model from \cite{Ven03} to derive a total gas column density of the order of N$_H$ $\sim$ 4.6 $\times$ 10$^{23}$ cm$^{-2}$ over the projected disk. Since the HF line is seen in absorption, the relevant H$_2$ column density, for the HF abundance calculation, is that in front of the continuum source, which should be $\sim$1/2 of the total hydrogen column density. Thus, we obtain a conservative lower limit of the HF abundance of $>$ 1.7 $\times$ 10$^{-9}$. Our observations suggest that, assuming a Galactic N(HF)/N$_H$ ratio, HF accounts for at least $\sim$10\% of the fluorine in the gas phase along the line of sight to the Cloverleaf quasar.


The spectrum shown in Fig. 2 shows an absorbing component shifted from the source systemic velocity by $\sim$-113 km s$^{-1}$. The CO $J=7-6$ line from \cite{Kne98} (their Figure 2), shows a marked asymmetry, with a steep rise on its blue side (-225 to -25 km s$^{-1}$) and a slower decrease on its red side (25 to 225 km s$^{-1}$). \cite{Kne98} found that the CO emission in the Cloverleaf originates in a disk- or ring-like structure. The absorbing component in our HF spectrum coincides in the velocity frame with excess emission (with respect 
to a standard Gaussian) on the blue side of the CO $J=7-6$ line. From the blue bump appearing on the CO $J=7-6$ and the velocity shifted HF line profile, we expect that the blue-shifted part of the CO and HF line arises from a region of the molecular torus which is positioned closer to the caustic of the lens. Future higher-sensitivity observations with ALMA are needed to establish the velocity distribution of HF and spatially resolve the four lensed lobes of the source.   

\acknowledgments

The authors would like to thank the CSO staff, especially to Brian Force, for his support during observations. The CSO is founded by the National Science Foundation under the contract AST-08388361. 

{\it Facilities:} \facility{CSO}, \facility{Herschel/HIFI}.

\clearpage

\begin{figure}
\centering
\includegraphics[angle=-90,scale=.70]{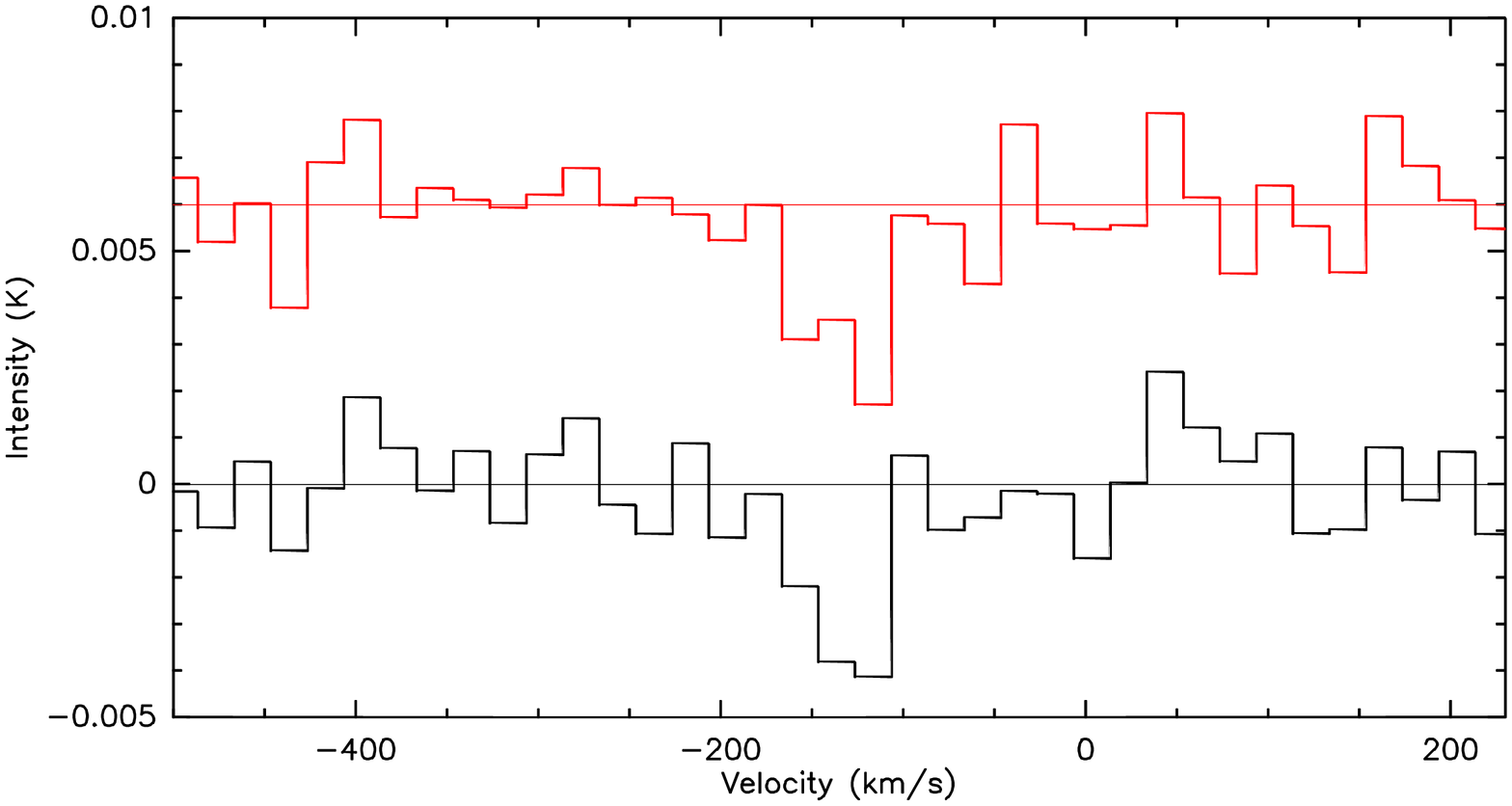}
\caption{Average spectra obtained with the FFTS (upper spectrum) and the AOS (lower spectrum) backends. Both spectra show an absorption feature at $\sim$-113 km s$^{-1}$.  \label{fig1}}
\end{figure}
\clearpage

\begin{figure}
\centering
\includegraphics[angle=-90,scale=.70]{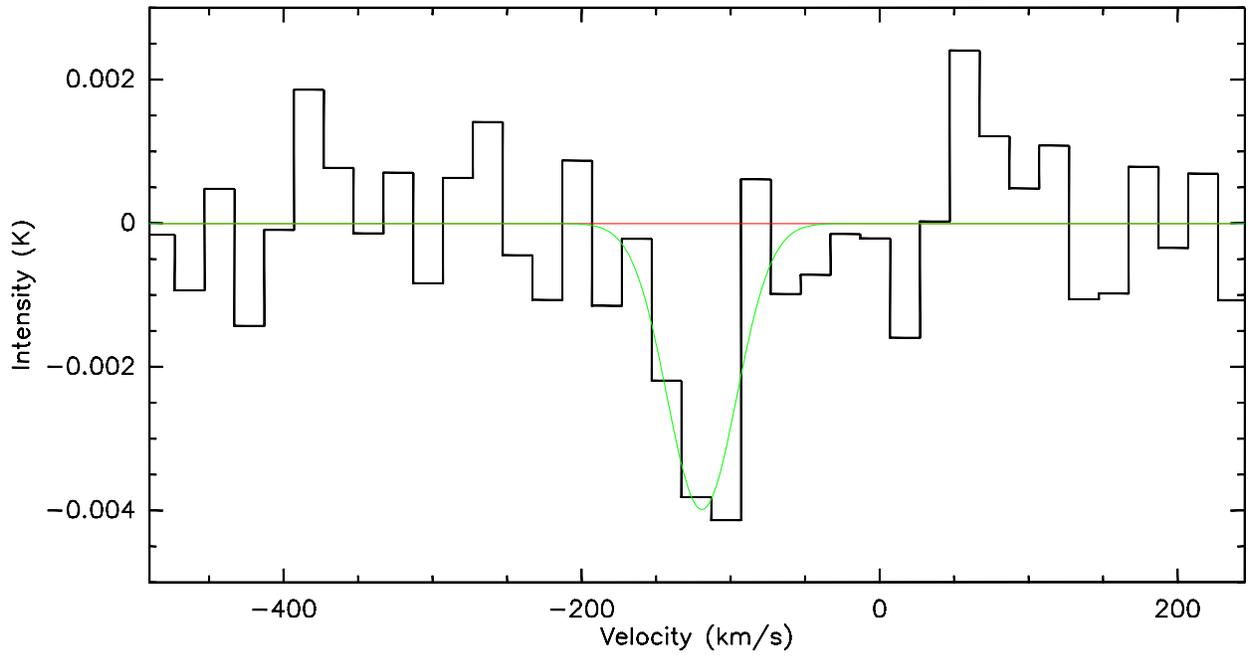}
\caption{Integrated spectrum of HF $J=1-0$ line toward the Cloverleaf quasar. The green thin curve shows a single-component Gaussian fit. \label{fig2}}
\end{figure}

\end{document}